\documentclass[%
reprint,
]{revtex4-2}

\usepackage{graphicx}
\usepackage{dcolumn}
\usepackage{bm}
\usepackage{hyperref}
\usepackage{multirow}

\begin{document}

\title{Stepping into the sea of instability: the new sub-$\mu$s superheavy nucleus $^{252}$Rf}

\author{J. Khuyagbaatar} 
\affiliation{GSI Helmholtzzentrum f\"ur Schwerionenforschung, 64291 Darmstadt, Germany}
\author{P. Mosat}
\affiliation{GSI Helmholtzzentrum f\"ur Schwerionenforschung, 64291 Darmstadt, Germany}
\author{J. Ballof}
\affiliation{GSI Helmholtzzentrum f\"ur Schwerionenforschung, 64291 Darmstadt, Germany}
\author{R.A. Cantemir}
\affiliation{GSI Helmholtzzentrum f\"ur Schwerionenforschung, 64291 Darmstadt, Germany}
\author{Ch.E. D\"ullmann}
\affiliation{GSI Helmholtzzentrum f\"ur Schwerionenforschung, 64291 Darmstadt, Germany}
\affiliation{Helmholtz Institute Mainz, 55099 Mainz, Germany}
\affiliation{Johannes Gutenberg-Universit\"at Mainz, 55099 Mainz, Germany}
\author{K. Hermainski}
\affiliation{GSI Helmholtzzentrum f\"ur Schwerionenforschung, 64291 Darmstadt, Germany}
\affiliation{Johannes Gutenberg-Universit\"at Mainz, 55099 Mainz, Germany}
\author{F.P. He\ss berger}
\affiliation{GSI Helmholtzzentrum f\"ur Schwerionenforschung, 64291 Darmstadt, Germany}
\author{ E. J\"ager}
\affiliation{GSI Helmholtzzentrum f\"ur Schwerionenforschung, 64291 Darmstadt, Germany}
\author{B. Kindler}
\affiliation{GSI Helmholtzzentrum f\"ur Schwerionenforschung, 64291 Darmstadt, Germany}
\author{J. Krier}
\affiliation{GSI Helmholtzzentrum f\"ur Schwerionenforschung, 64291 Darmstadt, Germany}
\author{N. Kurz}
\affiliation{GSI Helmholtzzentrum f\"ur Schwerionenforschung, 64291 Darmstadt, Germany}
\author{S. L\"ochner}
\affiliation{GSI Helmholtzzentrum f\"ur Schwerionenforschung, 64291 Darmstadt, Germany}
\author{B. Lommel}
\affiliation{GSI Helmholtzzentrum f\"ur Schwerionenforschung, 64291 Darmstadt, Germany}
\author{ B. Schausten}
\affiliation{GSI Helmholtzzentrum f\"ur Schwerionenforschung, 64291 Darmstadt, Germany}
\author{Y. Wei}
\affiliation{GSI Helmholtzzentrum f\"ur Schwerionenforschung, 64291 Darmstadt, Germany}
\affiliation{Johannes Gutenberg-Universit\"at Mainz, 55099 Mainz, Germany}
\author{P. Wieczorek}
\affiliation{GSI Helmholtzzentrum f\"ur Schwerionenforschung, 64291 Darmstadt, Germany}
\author{A. Yakushev}
\affiliation{GSI Helmholtzzentrum f\"ur Schwerionenforschung, 64291 Darmstadt, Germany}


\date{\today}

\begin{abstract}
	
We report the discovery of the new isotope $^{252}$Rf. With its extremely short half-life of $60^{+90}_{-30}$~ns, it expands the range of half-lives of the known superheavy nuclei by about two orders of magnitude. This nucleus was synthesized in its high-$K$ isomeric state, for which we measured a half-life of $13^{+4}_{-3}$~$\mu$s. Our results confirm a smooth onset of decreasing ground-state spontaneous fission half-lives in the neutron-deficient Rf isotopes towards the isotopic border of $10^{-14}$~s, which is the time needed to form an atomic shell. Our findings set a new benchmark for further exploration of phenomena associated with high-$K$ states and inverted fission-stability in the heaviest nuclei. 
\end{abstract}

\maketitle

The fission of heavy atomic nuclei is the main constraint on the existence of chemical elements with large atomic numbers. According to the semi-classical theory of fission, where the nucleus is considered as a charged nuclear liquid drop \cite{BohW39,HilW53a}, nuclei with proton numbers $Z\gtrsim103$ were expected to fission in less than $10^{-14}$~s, which, though, is the time needed for an atomic shell to be established \cite{Wap91}. On the other hand, the shell structure of the nucleus has a strong effect on the fission barrier, making superheavy nuclei (SHN) less fissile \cite{Str67a,Sobi66,Myers66,Nil69,Rand73,Baran86,Baran87,Xu,Delaroche06,Baran05,Kowal10,Moll15,Schunk16,Bender_2020}, which allows them to exist as the core of an atom. To date, this has been confirmed up to the element Og, with $Z=118$ \cite{HoM00,Oga17a,Hofmann17,Fritz17}. 
SHN with $Z\geq104$ and $N=149-177$ are known; their half-lives are in the range of $10^{-6}$--$10^{5}$~s \cite{Oga17a,nndc}. This supports the concept of increased stability in a sea of instability \cite{HoM00,Oga17a,Khu20c,Smits24}. Even the shortest measured half-lives are still much longer than $10^{-14}$~s, which precludes to precisely locate the isotopic borders in the region of the heaviest elements. Their locations, especially in the neutron-rich area, are crucial for nuclei with the largest $Z$ and $N$ produced in the astrophysical r-process \cite{Gor15a,Giu18,Cowan21}. Presently, such SHN relevant to the r-process are inaccessible experimentally.

In this regard, a recent discussion on the location of the isotopic border of SHN with $Z=104$ carried out in Ref.~\cite{Khu21b} has initiated an interesting debate. 
In Ref.~\cite{Khu21b}, the isotopic border of the neutron-deficient isotopes of Rf was experimentally shown to occur well below $N=146$, which promises further expansion, e.g., to $^{252}$Rf for which a spontaneous-fission half-life of $T_{\rm SF} \approx0.1$~$\mu$s was estimated. This value was in agreement with theoretical predictions, which gave half-lives of 0.65~$\mu$s \cite{Smol95} and $\approx$1~$\mu$s \cite{Khu22b} for the $T_{\rm SF}$ of $^{252}$Rf. On the other hand, in \cite{Lopez22}, recent experimental data were discussed and the half-life of $^{252}$Rf was concluded to be in the subpicosecond ($<1$~ps) range.  
Accordingly, the isotopic border for Rf was stated in \cite{Lopez22} to be almost reached, and $^{252}$Rf out of reach of current experimental capabilities. This scenario was initially challenged in Ref.~\cite{Khu21b}. Deviating conclusions between \cite{Khu21b} and \cite{Lopez22} have a significant impact on the understanding of the fission process in the heaviest nuclei \cite{Khu21b,Lopez22,Khu22b,Ackermann2024,Lopez23}. 

A direct synthesis of the hitherto unknown $^{252}$Rf would be an ideal solution for probing the above conclusions. However, the present experimental technique, i.e., in-flight separation of evaporation residues from fusion-evaporation reactions is not suited to identify nuclei with half-lives shorter than about 10$^{-6}$~s, which roughly is the flight time through the separator. Another approach, suitable for identifying shorter-lived nuclei is to imply a longer-lived isomeric state \cite{Baran86,Baran87,Xu,Herzberg2006,Adamian10,Walker_2012,WalX16,Khu22b,Ackermann2024}. Such cases occur in $^{270}$Ds \cite{270Ds_Hof}, $^{250}$No \cite{Belozerov2003,Peter06,Svir17,250No_shels,Kal20a} and $^{254}$Rf \cite{DavC15a}, where multi-quasiparticle (qp) states were found that are longer-lived than the ground states. Therefore, the identification of $^{252}$Rf via a high-$K$ state has been discussed in Refs.~\cite{Adamian10,Khu22b}. In both works, the theoretically \cite{Liu} predicted low-lying high-$K=6^+$ state in $^{252}$Rf was considered. In Ref.~\cite{Khu22b} its fission half-life was estimated to be long enough (50~$\mu$s) for experimental detection.

In this letter, we report on the synthesis of the new isotope $^{252}$Rf, and the observation of both its ground state and a high $K$-isomeric state. 

The experiments were carried out at GSI Helmholtzzentrum f\"ur Schwerionenforschung, Darmstadt, Germany. A $^{50}$Ti$^{12+}$ beam (5\,ms-long pulses with 5~Hz repetition rate) was accelerated by the UNIversal LInear ACcelerator (UNILAC). Targets of $^{204}$PbS (isotopic composition: 99.94\% $^{204}$Pb; 0.04\% $^{206}$Pb; 0.01\% $^{207}$Pb; 0.01\% $^{208}$Pb) with an average thickness of (0.5-0.8)~mg/cm$^{2}$ were used. They were mounted on a wheel which was rotating synchronously with the beam pulse structure \cite{JaeB14a}. The irradiation was performed at four different beam energies of $E_{\rm b}=239.1$, 241.4, 241.9 and 244.1~MeV in the middle of the target. These energies were chosen to cover the region of the maximum of the excitation function of the 2n channel of the $^{50}$Ti~+~$^{204}$Pb fusion-evaporation reaction as calculated by the statistical code HIVAP \cite{Rei81} with parameters adjusted to describe the experimental data of the $^{50}$Ti~+~$^{206-208}$Pb reactions \cite{Fritz97,Khu20a,Khu21a} and the 1n channel of the $^{50}$Ti~+~$^{204}$Pb \cite{Fritz97,Khu21b,Lopez22}. Nuclear masses were taken from the AME2020 \cite{AME2020} and those missing in \cite{AME2020} were taken from the FRDM \cite{Moll95}.

Separation and collection of evaporation residues (ERs) was performed in the gas-filled recoil separator TASCA \cite{SemB08a}, which was operated with helium gas at 0.8\,mbar pressure and with a magnetic rigidity ($B\rho$) of 2.14~Tm \cite{Khu12a,Khu13b}. The efficiency of TASCA to guide ERs to the focal plane detector was estimated to be 60\% \cite{Gre13a,Khu12a,Khu20a}. A double-sided silicon detector with 144 vertical ($X$) and 48 horizontal ($Y$) strips on the front and back sides, respectively, was used to  to register the implantation and decay of the ER. Signals from the $X$ and $Y$ strips were pre-amplified with different gains to provide two energy branches up to about 20 and 200~MeV, respectively.
The data acquisition system was triggered by both $X$ and $Y$ strips. However, only 118 $X$ strips were read out. All pre-amplifier signals were digitized by 100~MHz-sampling FEBEX4 analog-to-digital converters \cite{Kurz12,Febex4}. About 95\% of the data were taken with 80~$\mu$s-long traces and the rest with 60~$\mu$s. The pretrigger time was always 4.7~$\mu$s \cite{Khu22a}. Energy resolutions (FWHM) of both $X$ and $Y$ strips were about $50$~keV for 5.8-MeV $\alpha$ particles from an external $^{244}$Cm source. Energy calibrations were made using $\alpha$ decays of nuclei produced in transfer channels with $^{206,207,208}$Pb targets. The average beam intensity on the targets was $\approx4\times 10^{11}$~s$^{-1}$, which resulted in an average counting rate of about 15~s$^{-1}$. For more details on the TASCA detection system and data analysis see Refs.~\cite{Khu19b,Khu20b,Khu21a,Khu22a}.

\begin{figure}[t]
	\vspace*{-0mm}
	\hspace*{-3mm}
	\resizebox{0.495\textwidth}{!}{\includegraphics[angle=0]{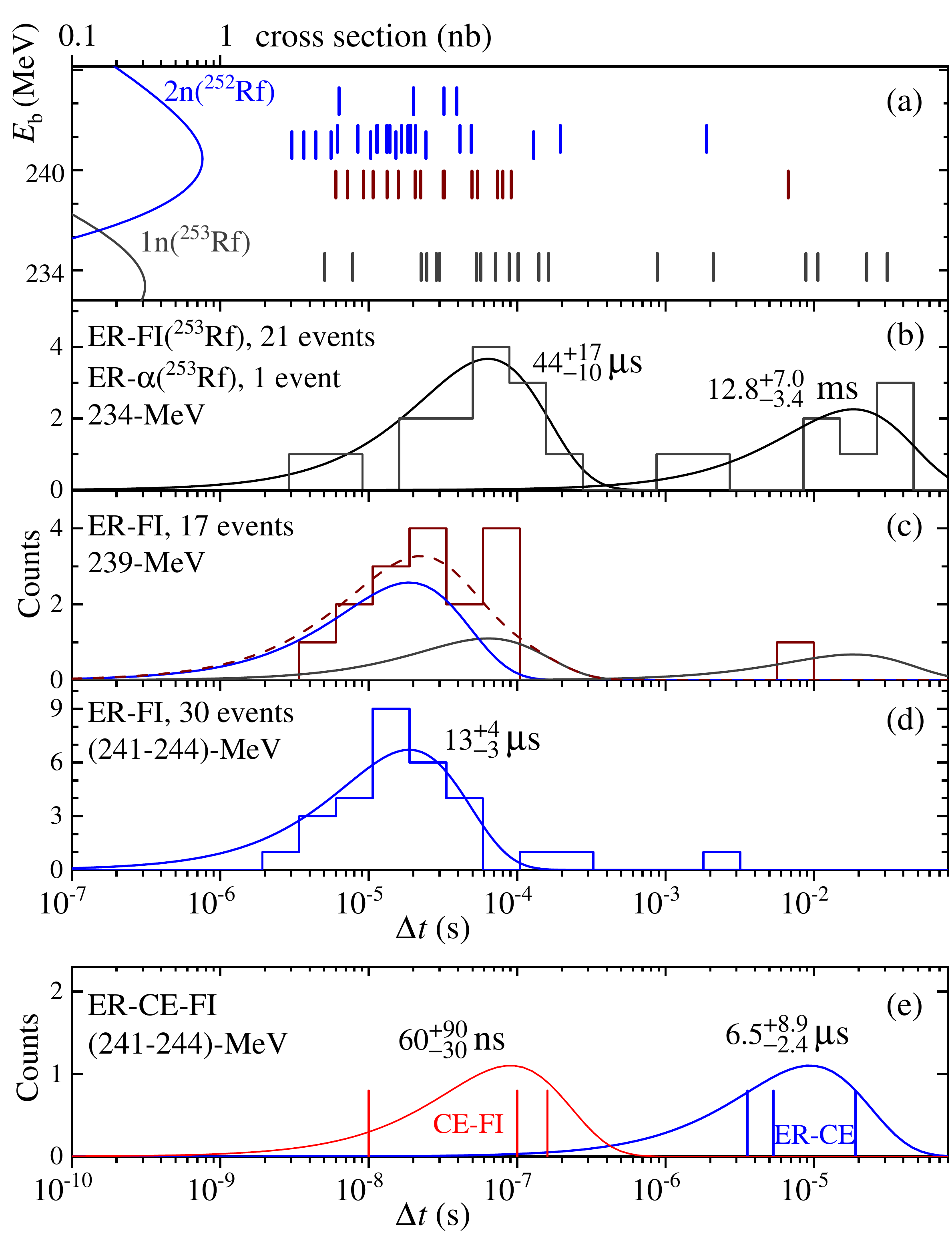}}
	\vspace*{-6mm}
	\caption{\label{lifetime} (a) Individual correlation times of the measured ER--FI events are shown as a function of the beam energy. In total 17, 8, 18, and 4 events measured at $E_{\rm b}\approx239$, 241, 242, and 244~MeV, respectively, are shown. Not all events can visually be distinguished as the $\Delta t$ values of some events are similar. In addition, the 21 ER--FI events measured at 234~MeV taken from Ref.~\cite{Khu21b} are shown. The 1n and 2n excitation functions for the $^{50}$Ti~+~$^{204}$Pb reaction calculated with HIVAP (see text) are also shown. (b-d) The time distributions of ER--FI events observed at 234, 239 and (241-244)~MeV. The time distribution at 234~MeV includes one additional ER--$\alpha$ event with $\Delta t$=40.8~ms observed in \cite{Khu21b} and assigned there to $^{253}$Rf. (e) Times of the CE and FI events from the three ER--CE--FI correlations calculated relative to the preceding ER and CE events, respectively, are shown. For the CE--FI sequences, deadtime-corrected correlation times ($\Delta t - 90~$ns) are shown. Curves represent the calculated density distribution of the events on a logarithmic time scale according to Ref. \cite{SchS84a}. For details see text.}
\end{figure} 

\begin{figure*}[t]
	\vspace*{-4mm}
	\centering
	\hspace*{-2 mm}
	\resizebox{1.00\textwidth}{!}{\includegraphics[angle=0]{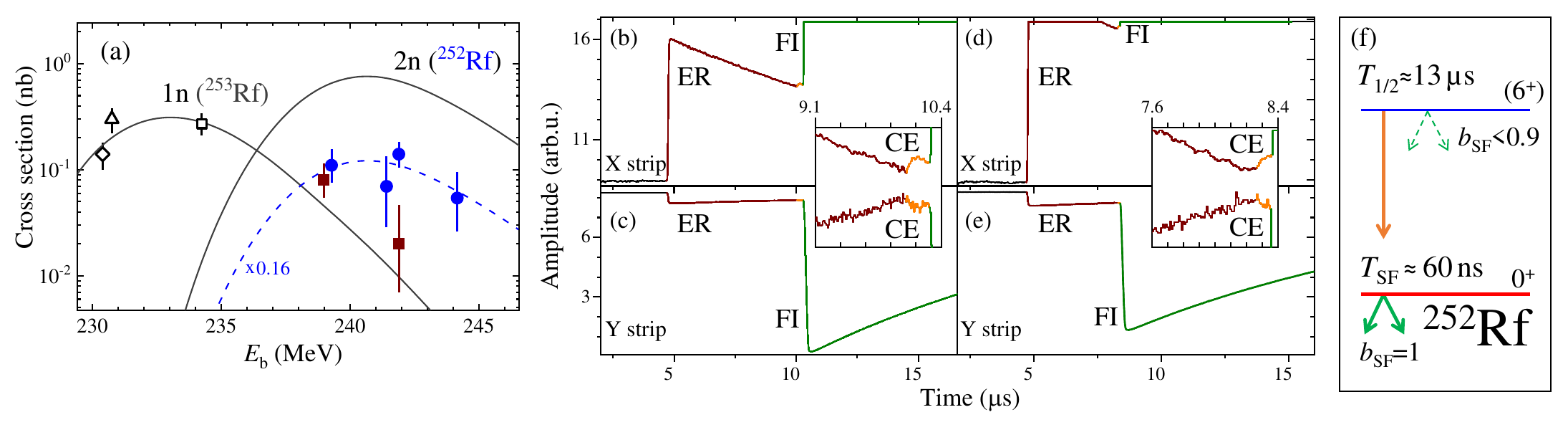}}
	\vspace*{-4mm}
	\caption{(Color online) (a) The experimental and calculated (HIVAP, see text) cross sections of the $^{50}$Ti~+~$^{204}$Pb reactions are shown by symbols and lines, respectively. The open and solid symbols show the literature 1n data (triangle \cite{Fritz97}, squares \cite{Khu21b} and diamond \cite{Lopez22}) and the present (squares: 1n, and circles: 2n) data, respectively. The dashed line corresponds to the 2n cross sections reduced by a factor of 0.16. Two ER-traces are shown in (b,c) and (d,e) as examples of three ER--CE--FI events. Insets show a closer view of the traces in the time regions, where the CE was detected. (f) The proposed decay schemes for the isomeric and ground states of $^{252}$Rf. See text for details.
	}
	\label{figure2}
\end{figure*}

At $E_{\rm b}\approx239$, 241, 242, and 244~MeV, we identified 17, 8, 18, and 4 spatially (same $X$ and $Y$ strips) correlated events, respectively, each consisting of ER and fission (FI) signals with correlation times ($\Delta t$) up to 10~ms. For most ER--FI events, both signals were identified in the same trace. 

Individual $\Delta t$ values of all ER--FI events are shown in Fig.~\ref{lifetime}(a) as a function of $E_{\rm b}$. In addition, the ER--FI events, which were previously measured at 234~MeV  \cite{Khu21b}, are shown. Moreover, to present the influence of beam energy on the production of the 1n and 2n channels of the $^{50}$Ti~+~$^{204}$Pb reaction, the calculated excitation functions are also shown. In this $E_{\rm b}$--$\Delta t$ plot, the region of highest density of ER--FI events is shifting towards shorter times with increasing $E_{\rm b}$. 

For quantitative analysis, the time distributions of ER--FI events measured at $E_{\rm b}\approx234$, 239, (241-244)~MeV are shown in Figs.~\ref{lifetime}(b-d). The $\Delta t$ of one ER-$\alpha$ event from $^{253}$Rf observed in \cite{Khu21b} was also included in panel (b).

As presented in Ref.~\cite{Khu21b}, the ER--FI events at 234~MeV were unambiguously attributed to $^{253}$Rf. The time distribution (Fig.~\ref{lifetime}b) presents the decays from two states in $^{253}$Rf with half-lives of $44^{+17}_{-10}$~$\mu$s and $12.8^{+7.0}_{-5.4}$~ms, which were recently confirmed in Ref.~\cite{Lopez22}. Fission events from $^{253}$Rf, dominant at this lowest energy $E_{\rm b}$, are important for the analysis of the presently obtained ER--FI data at higher $E_{\rm b}$. According to the excitation functions (see Fig.~\ref{lifetime}a), the total set of 30 ER--FI events measured at (241-244)~MeV should be minimally contaminated by decays of $^{253}$Rf. The time distribution of these ER--FI events shown in Fig.~\ref{lifetime}(d) results in a peak, which is well described by radioactive decay with a half-life of $13^{+4}_{-3}$~$\mu$s according to the procedure suggested in \cite{SchS84a}. One the basis of the fitted time distribution, the three ER--FI events with the longest $\Delta t$ (Fig.~\ref{lifetime}d) are likely from $^{253}$Rf. Among the remaining 27 ER--FI events with a half-life of $13^{+4}_{-3}$~$\mu$s, we do not exclude a contamination of about two events from the 44-$\mu$s decay of $^{253}$Rf. These two events are within the statistical uncertainty of 27(5) events, and thus not significant. 

In the case of the intermediate $E_{\rm b}$ of 239~MeV, the time distribution of the 17 ER--FI events should then be a superposition of the three fission activities with half-lives of 13~$\mu$s, 44~$\mu$s, and 12.8~ms. This is fully supported by the broad time distribution shown in Fig.~\ref{lifetime}(c). The data were disentangled into the three components as also shown in Fig.~\ref{lifetime}c). Finally, on the basis of these time analyses, the hitherto unknown fission activity with a half-life of $13^{+4}_{-3}$~$\mu$s observed at $E_{\rm b}$=(241-244)~MeV is attributed to originate from $^{252}$Rf. 

The cross sections deduced from the numbers of ER--FI events attributed to $^{252}$Rf and $^{253}$Rf are shown in Fig.~\ref{figure2}a) together with values measured for the 1n channel in \cite{Lopez22} and \cite{Fritz97}. These values are compared with our excitation functions calculated by HIVAP \cite{Rei81}.

The cross sections deduced for the 13-$\mu$s fission activity are smaller than the predictions from HIVAP, but their trend agrees well with the predicted excitation function of the 2n channel. Cross-sections smaller than expected could be due an overestimation by HIVAP and/or could indicate a population of a $K$-isomeric state in $^{252}$Rf as expected \cite{Adamian10,Khu22b}. Moreover, the 13~$\mu$s half-life is anomalously long for the ground-state $T_{\rm SF}$ of $^{252}$Rf \cite{Smol95,Khu21b,Khu22b,Lopez22}. On the other hand, a value of 13~$\mu$s is in the range expected for a $K$-isomeric state \cite{Khu22b}. Such isomeric states are prevalent in the region of Fm-Rf isotopes. Typically, population probabilities of low-lying 2 quasiparticle (qp) states in fusion-evaporation reactions are $\geq$10\% \cite{252No_Sulig,254No_Hess,DavC15a,Kal20a,Khu20a,Kal20a,Khu21a}. Assuming the 13 µs fission activity might represent the decay of an isomer, we therefore searched for signatures for the de-excitation of a $K$-isomeric state in the data.

The de-excitation of high-$K$ states is often associated with the emission of multiple conversion electrons (CEs), which are typically used for their identification and study \cite{JONES2002471,254no_Tandel,AckT17,Ackermann2024}.   
In the last decade, the implementation of fast digital sampling analog-to-digital converters in superheavy element research greatly advanced the efficient registration of CEs from high-$K$ states. This so-called triggerless CE measurement \cite{Khu20a} enables the safe and highly-efficient identification of CEs, i.e., the decay of isomeric states produced with a small number (down to a few) of events \cite{Khu21a,Khu22a,Khu20b,Khu21b}. Therefore, among the ER--FI events measured at (241-244)~MeV and attributed to the decay of a $K$-isomeric state in $^{252}$Rf, we searched for a third, small energy CE signal detected in between the ER and FI signals. 

To clearly identify a small-energy signal from a CE preceding a large FI signal, a minimum time difference of $\approx$90~ns ($\approx$9 samples) is necessary. This is thus the deadtime of our system, and precludes the identification of CE signals occurring less than 90~ns before a FI event. A detailed trace-inspection of all 30 ER--FI correlations event-by-event revealed three such ER--CE--FI sequences. In each case, all three signals were measured in a single ER trace; the CEs were detected shortly (100, 190 and 250~ns) before the FI signals. Two such ER traces measured in both the low-energy ($X$-strip) and high-energy ($Y$-strip) branches are shown in Fig.~\ref{figure2}(b-e). These represent the cases, where the CE was detected with $\Delta t=250$ and 100~ns, respectively, relative to the FI signal. 

In Fig.~\ref{lifetime}(e), $\Delta t$ between the ER--CE of the three ER--CE--FI events are shown. They lead to a half-life of $6.5^{+8.9}_{-2.4}$~$\mu$s, which matches the 13-$\mu$s half-life of the $K$-isomeric state. The energies of the three CEs were $\approx$0.22, $\approx$0.28 and $\approx$0.32~MeV. This points at an excitation energy of the $K$-isomeric state of $\geq0.2$~MeV. Typically, energies of most CE signals from the decay of $K$-isomeric states represent only a fraction of an initially high excitation energy ($\approx1$~MeV) because they originate from highly-converted excited states populated by a $\gamma$ transition. 

Therefore, the ER--CE events are attributed to the implantation of $^{252}$Rf in its $K$-isomeric state, which is followed by its electromagnetic decay into the ground state. To calculate the ground-state spontaneous-fission half-life from the three CE--FI events, a 90~ns deadtime was subtracted from each $\Delta t$, resulting in residual times of 10, 100, and 160~ns as also shown in Fig.~\ref{lifetime}(e). From these three events, a half-life of $T_{\rm SF}=60^{+90}_{-30}$~ns was deduced. The $^{252}$Rf directly populated in its 60-ns ground state has a very small chance to survive the $\approx$0.6~$\mu$s flight through TASCA, which explains the non-observation of any ER--FI event with $\Delta t$ in the sub-$\mu$s region (see Fig.~\ref{lifetime}(a)) and the fact that the measured 2n cross-sections deduced from the $13$~$\mu$s state with a maximum value of $0.14^{+0.5}_{-0.4}$~nb at $E_{\rm b}$=241.9~MeV are lower than predicted by HIVAP, which includes all 2n channel outcomes.     

Our data do not allow determining the spin and parity of the observed $K$-isomeric state. Theory predicts the occurrence of several low-lying 2qp high-$K$ states in $^{252}$Rf \cite{Liu,Adamian10}. For instance, according to Ref.~\cite{Liu} low-lying 2qp high-$K$ states with $\nu^2$6$^+$, $\nu^2$8$^-$, $\pi^2$5$^+$ in $^{252}$Rf are predicted to be located at excitation energies of 0.88, $\approx$1.1, and $\approx$1.1~MeV, respectively. In fact, many theoretical works predict $\nu^2$6$^+$ as the lowest-lying qp state in the $N=148$ isotones ($^{252}$Rf, $^{250}$No, $^{248}$Fm, $^{246}$Cf, and $^{244}$Cm) \cite{Delaroche06,Liu11,Adamian10,Walker_2012,Minkov22}. In the cases of known $K$-isomers in $^{244}$Cm \cite{HANSEN1963410} and $^{250}$No \cite{250No_shels}, the experimental data confirm the theoretically predicted spin and parity of 6$^+$, while for $^{248}$Fm only a tentative assignment is given \cite{Kondev15}. Thus, based on these results, we propose tentatively 6$^+$ for the spin and parity of the presently observed 13-$\mu$s state. 

We note that recently in $^{250}$No, the signature for a second, higher-lying isomeric state populated with a probability of $<1$\%, was observed \cite{Khu22a}. Such a low value is typical for 4qp states: all known cases in Fm-Rf feature values $<5$\% while the lower-lying 2qp states $\geq 10$\%. The presently observed $K$-isomer is thus unlikely to originate from a 4qp state because the negligibly low production rate of the latter that would imply the total 2n cross-section values much larger than HIVAP. Since the parameters of HIVAP adjusted to fit the 1n cross-sections for the production of $^{253}$Rf, which is more stable against fission than $^{252}$Rf an increase in 2n cross sections is unlikely. Our suggested decay scheme for $^{252}$Rf is sketched in Fig.~\ref{figure2}(f). Based on the numbers of ER--FI (24) and ER--CE--FI (3) events assigned to $^{252m}$Rf, a value of 0.9 is given as an upper limit value for the fission branching.

\begin{figure}[t]
	\hspace*{-4mm}
	\includegraphics[width=0.5\textwidth]{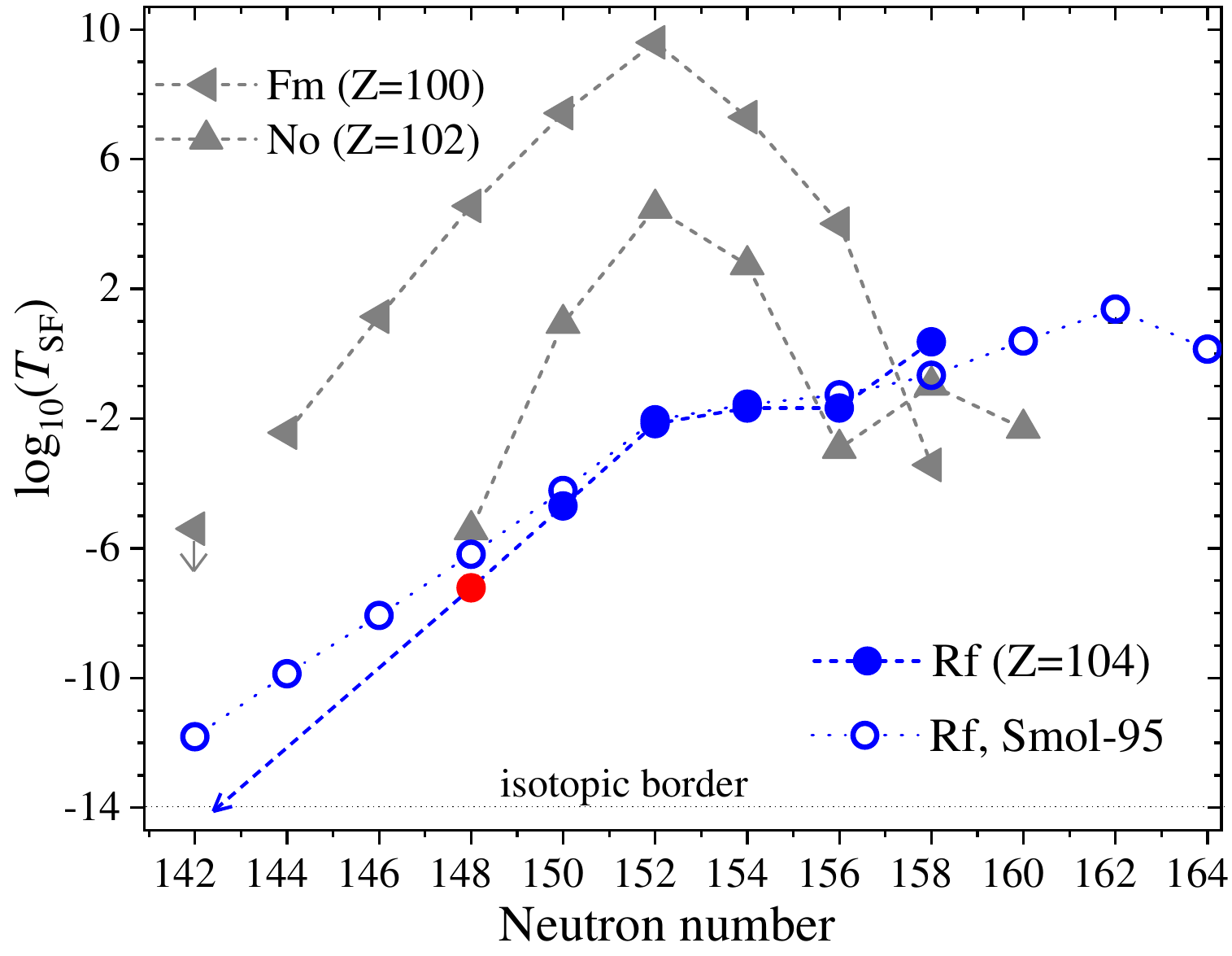}
	\vspace*{-6mm}
	\caption{\label{Tsf} Spontaneous fission half-lives of Fm, No, and Rf isotopes \cite{Fritz17} are shown by solid symbols. Theoretically predicted half-lives for Rf (Smol-95, \cite{Smol95}) are shown by open circles. The presently measured value for $^{252}$Rf is highlighted. The horizontal dotted line at $10^{-14}$~s indicates an isotopic border for the existence of the chemical elements. See text for details.}
\end{figure}

Our measured SF half-life of $^{252}$Rf is shown in Fig.~\ref{Tsf} in comparison with experimental data for other Fm-Rf isotopes. The systematics of $T_{\rm SF}$ for Fm and No shows the well-pronounced stabilization effect of the shell gap at $N=152$. However, the $T_{\rm SF}$-trend of the Rf isotopes does not show an analogue peak, i.e., $T_{\rm SF}$ of $^{256}$Rf is short. This feature, known for decades \cite{Oga75} and initially predicted by theory \cite{Nil69}, was explained as an effect of the disappearance of the outer barrier of the double-humped fission barrier (as it is typical for Fm and No isotopes) in $^{256}$Rf. Accordingly, in more neutron-deficient Rf isotopes, theory predicts no further sudden drop in $T_{\rm SF}$ since their single-humped fission-barrier heights decrease smoothly \cite{Nil69,Rand73,HESSBERGER1986445,Baran05,Delaroche06,Ward12,Kowal10,Moll15,Schunk16,Khu20c}. In Fig.~\ref{Tsf}, predicted $T_{\rm SF}$ for the Rf isotopes, taken from Ref.~\cite{Smol95} are shown as a representative of theory \cite{Stas89,Smol95,Ward12,Schunk16,Bender_2020}. The present experimental $T_{\rm SF}$($^{252}$Rf) is in line with the theoretical prediction \cite{Smol95}. 

The expanded systematics of $T_{\rm SF}$ in Fig.~\ref{Tsf}, still shows a stabilizing effect of the shell gap at $N=152$ on the fission-stability of the Rf isotopes, albeit ``weaker'' than those in Fm and No. This is attributed to the impact of the single-humped nature of the fission barriers in Rf isotopes \cite{Khu20c}, a weakening of the shell effect at $N=152$ in Rf, and/or a stabilizing effect originating from the next enhanced shell gap at $N=162$ \cite{Sobi07,Khu16a}. Theoretical $T_{\rm SF}$-values, where the effects of both $N=152$ and $N=162$ on the fission stability are considered describe the experimental data well (see Fig.~\ref{Tsf}). This shows a crucial impact of the enhanced shell gap on the fission-stability of SHN. An extrapolation of the experimental data suggests that the isotopic border in Rf will be crossed by $N=142$ while theory \cite{Smol95} predicts $N<142$. An examination of the $N=152$-effect on the fission ($T_{\rm SF}$) of more neutron-deficient heavier SHN, e.g., in Sg ($Z=106$) is essential to explore the evolution of the isotopic borders.

In conclusion, we identified the new isotope $^{252}$Rf and have observed its ground-state decay as well as the decay of a high-$K$ isomeric state with half-lives of $60^{+90}_{-30}$~ns and $13^{+4}_{-3}$~$\mu$s, respectively. The ground-state half-life of $^{252}$Rf pushes the limit of known fission half-lives of SHN down by about two orders of magnitude and shows that the isotopic border for the neutron-deficient Rf isotopes is yet to be reached. The present findings show that the inversion of fission stability between the high-$K$ and the ground states is likely a widespread feature in the neutron-deficient superheavy nuclei as was suggested in Ref.~\cite{Khu22b}. High-$K$ phenomena should thus be explored further in heavier nuclei. 

We are grateful to GSI's ion-source and UNILAC staff, the Experiment Electronics, and the Target Laboratory departments for their support of the experiment. This work was supported by the German BMBF (Grant Number 05P21UMFN2). The results presented here are based on the experiments U328 and G-22-00034, which were performed at the beam line X8/TASCA at the GSI Helmholtzzentrum f\"ur Schwerionenforschung, Darmstadt (Germany) in the frame of FAIR Phase-0. 


\bibliography{references_2024.bib}

\end{document}